\documentclass[12pt]{article}
\usepackage{epsfig}
\def\nin{\rm{N_{i}}}
\def\nhid{\rm{N_{h}}}
\def\ntr{\rm{N_{t}}}
\def\lr{\rm{LR}}
\def\npe{\rm{N _{pe}}}
\def\leg{\rm{D _{\gamma}}}
\def\lenr{\rm{D _{nr}}}

\def\e90{\rm{\epsilon _{90} }}
\def\l90{\rm{\lambda _{90} }}

\begin{document}
\hfill AS-TEXONO/03-05 \\
\hspace*{1cm} \hfill \today

\begin{center}
{\Large  \bf
Near Threshold
Pulse Shape Discrimination Techniques in 
Scintillating CsI(Tl) Crystals
}\\
\vskip 0.5cm
\large
S.C.~Wu$^{a,b}$,
Q.~Yue$^{c,d}$,
W.P.~Lai$^{a,e}$,
H.B.~Li$^{a,b}$,
J.~Li$^{c,d}$,
S.T.~Lin$^{a,b}$,
Y.~Liu$^{d}$,\\
V.~Singh$^{a}$,
M.Z.~Wang$^{b}$,
H.T.~Wong$^{a,}$\footnote{Corresponding~author:
Email:~htwong@phys.sinica.edu.tw;
Tel:+886-2-2789-6789;
FAX:+886-2-2788-9828.},
B.~Xin$^{f}$,
Z.Y.~Zhou$^{f}$

\normalsize
\vskip .2cm

\begin{flushleft}
{$^{a}$\rm 
Institute of Physics, Academia Sinica, Taipei 115, Taiwan.\\}
{$^{b}$\rm
Department of Physics, National Taiwan University, Taipei 106, Taiwan.\\}
{$^{c}$\rm 
Department of Engineering Physics, Tsing Hua University, 
Beijing 100084, China.\\}
{$^{d}$\rm
Institute of High Energy Physics, Beijing 100039, China.\\}
{$^{e}$\rm 
Department of Management Information Systems,
Chung Kuo Institute of Technology, \\
\hspace*{1cm} Hsin-Chu 303, Taiwan.\\
{$^{f}$\rm 
Department of Nuclear Physics, Institute of Atomic Energy, 
Beijing 102413, China.\\}
}
\end{flushleft}

\end{center}
\vskip 0.5cm
\begin{abstract}

There are recent interests with
CsI(Tl) scintillating crystals for Dark Matter
experiments.
The key merit is the capability
to differentiate nuclear recoil (nr) signatures
from the background $\beta / \gamma$-events
due to ambient radioactivity on
the basis of their different pulse shapes.
One of the major experimental challenges is 
to perform such pulse shape analysis in the 
statistics-limited domain 
where the light output is close to the detection 
threshold.
Using data derived from 
measurements with low energy $\gamma$'s
and nuclear recoils due to neutron elastic scatterings,
it was verified that the pulse shapes between
$\beta / \gamma$-events are different.
Several methods of pulse shape discrimination
are studied, and their relative merits 
are compared.
Full digitization of the pulse shapes is crucial
to achieve good discrimination.
Advanced software techniques with mean time, 
neural network and likelihood ratios give rise to
satisfactory performance, and are superior to the 
conventional Double Charge method 
commonly applied at higher energies. 
Pulse shape discrimination becomes effective 
starting at a light yield of 
about 20 photo-electrons.
This corresponds to a detection threshold 
of about 5~keV electron-equivalence energy,
or 40$-$50~keV recoil kinetic energy, 
in realistic experiments.

\end{abstract}

\begin{flushleft}
\end{flushleft}

\vfill


\newpage

\section{Introduction}

The detection of Dark Matter and the studies of their
properties~\cite{pdg} are of fundamental importance in
particle physics and cosmology.
The Weakly Interacting Massive Particles (WIMPs)
are good candidates for ``Cold Dark Matter'',
and their experimental searches
have gathered a lot of interests in recent years. 
The most promising avenue is to detect the
nuclear recoil signatures due to elastic scatterings
of WIMPs on the target isotopes.
The typical energy depositions are
only of the order of 10~keV, imposing
big experimental challenges in terms
of the detection of weak signals as well as 
background control at low energy close to detection
threshold.
A wide spectrum of experimental techniques is 
being pursued~\cite{pdg}.
There is still much room for new detector concept 
to push the sensitivities further. It would be
of great interest if the sensitivities of 
WIMP searches can probe the level predicted by
the various Super-Symmetry models.

There are potential merits of using CsI(Tl) scintillating
crystals~\cite{csichar} for WIMP search and other low-energy
low-background experiments~\cite{prospects,ksexpt}.
An experiment with 200~kg of CsI(Tl) crystal
scintillators to study low energy neutrino
interactions at the Kuo-Sheng power reactor is
being pursued~\cite{ksexpt,proto}, while
the adaptation of the crystal
for Dark Matter searches
is the focus of several R\&D
projects~\cite{csidmfrance,csidmuk,qfpaper}
and an on-going experiment~\cite{csidmkorea}.

The high-A content of the CsI enhances
the sensitivities for the spin-independent interactions
(which depends on the neutron number squared)
between the WIMPs and the target,
relative to most other candidate target isotopes.
The high-Z composition allows
a compact design and provides large suppression of
background due to ambient radioactivity if
a three dimensional fiducial volume definition
can be realized.
Both $^{133}$Cs and $^{127}$I are 100\% in their
respective isotopic abundance. Being close in
their mass numbers, the response to nuclear recoil
from the interactions with WIMPs would be similar,
allowing simpler interpretations of 
the experimental signatures.

As a detector, the crystal  
has large light yield, low energy
threshold and with pulse shape
discrimination (PSD) characteristics for 
differentiating $\beta$/$\gamma$ background
from the nuclear recoil events~\cite{csichar,proto}.
Scintillating NaI(Tl) crystals with
the order of 100~kg target mass have been
deployed for Dark Matter experiments~\cite{dama},
but it has been shown that CsI(Tl) provides
superior PSD capabilities to NaI(Tl)~\cite{csidmfrance,csidmkorea}.
Unlike NaI(Tl), CsI(Tl) is only
slightly hygroscopic such that it
can be machined easily and does not
require hermetic seal (that is, passive materials)
in a large detector system. 
In addition, large (40~tons) electromagnetic
calorimeter systems~\cite{bfactories}
have been constructed and
made operational in high energy physics experiments,
making this technology affordable and
realistic to scale up.
Considering all the associated costs, the price
of CsI(Tl) is in fact less than that for NaI(Tl).
In  order to produce positive and definite evidence
of the WIMPs, 
an accurate measurement of the annual modulation
(where the maximal effects are only 7\%)
would be necessary such that
the availability of large target mass 
is a very desirable feature.

One of the key issues to realize a Dark Matter search
experiment with CsI(Tl) crystal scintillator
is the studies of the experimental signatures 
of nuclear recoils due to WIMP-nuclei elastic
scatterings. Nuclear recoils produce 
high charge density (DA/dx)
such that the scintillating light yield is ``quenched''
and the timing profile of pulse shape is different
relative to the same energy deposition by minimum ionizing
particles~\cite{scinbasic}. 
These WIMP-induced
signatures are the same as the nuclear recoil
events produced by elastic scattering of neutrons
on nuclei, and hence can be studied in the
laboratory. 


\section{Pulse Shape Discrimination}

It has been well-studied~\cite{csichar,proto} 
that the light emission profiles of 
scintillating CsI(Tl) crystals
exhibit different shape for
$\gamma$-rays and electrons (that is, minimum ionizing particles),
as compared to that for
$\alpha$-particles and nuclear recoils at
the high energy ($>$100~keV) regime~.
Heavily ionizing
events  due to $\alpha$-particles and nuclear
recoils have
{\it faster} decays than those from e/$\gamma$'s $-$
opposite to the response in
liquid scintillator~\cite{scinbasic}.
This characteristic property makes particle identification
possible with this scintillator~\cite{psdpid}.

Matured pulse shape discrimination (PSD) 
techniques have been devised
at high energies where the photo-electrons
are abundant. The experimental challenge for
adapting the PSD idea to Dark Matter experiments
is that one must now work in the
regime where the number of photo-electrons ($\npe$)
is small such that the statistical fluctuations
may wash out the differences.

In the following sub-sections,
we verify that the pulse
shapes of $\gamma$ and nuclear recoil
events with CsI(Tl) crystals
are different even at the
Dark Matter relevant low energy
regime ($<$100~keV).
The theme and focus of this article
is to investigate and compare the
various software techniques
which can perform {\it event-by-event}
PSD at this low light output domain.
Detailed response characteristics of 
CsI(Tl) crystals at the low energy regime
have already been studied in previous 
work~\cite{csidmfrance,csidmuk,qfpaper,csidmkorea}.

\subsection{Measurements}

A CsI(Tl) crystal of dimensions 
$\rm{5~cm \times 5~cm \times 5~cm}$
and mass 560~g
was used to provide data for these investigations.
The light emissions were read out
by a 29~mm diameter photo-multiplier 
tube (PMT)\footnote{CR110, Hamamatsu Photonics, China}
with standard bi-alkali photo-cathode.
The conversion factor
between energy deposition and light output is
4 photo-electron per keV of electron-equivalence (e.e.) energy.
This was obtained by calibration measurements
with an LED pulser operated at
the  $\npe \sim 1$ intensity.
The events were digitized
by a 20~MHz (that is, 50~ns for one time-bin)
Flash Analog-to-Digital Convertor 
(FADC)~\cite{electronics} with 8-bit resolution,
such that the pulse shape can be denoted
by an amplitude time-sequence $\rm{A_i}$.

The low energy $\gamma$-data were taken with
a standard radioactive $^{133}$Ba source,
which provides several $\gamma$-lines up
to 356~keV. The series of associated
$\gamma$-peaks provide good energy
calibration at low energy.
The low energy ($<$30~keV) events crucial
for this study are
due to Compton scatterings of the higher
energy $\gamma$'s such that they originate
from the bulk of the crystal.
These merits justify the choice of $^{133}$Ba
source over other low energy sources like
$^{55}$Fe (5.9~keV) and $^{109}$Cd (22.1~keV)
where the attenuation of the $\gamma$'s due to
the crystal wrapping materials is severe, and the
events only originate at the surface of the
crystal. 
Nuclear recoil data, on the other hand,
were taken from the neutron facility at 
the 13~MV Tandem accelerator
at the China Institute of Atomic Energy at Beijing.
The data consisted of
Time-of-Flight (ToF) measurements which
helped to distinguish nuclear recoil
from the other background events.
The results of the quenching factor measurements
were already published~\cite{qfpaper}.
For completeness,  $\alpha$-events
with $^{241}$Am (5.49~MeV) source were also recorded.

\begin{figure}
\centerline{
\epsfig{file=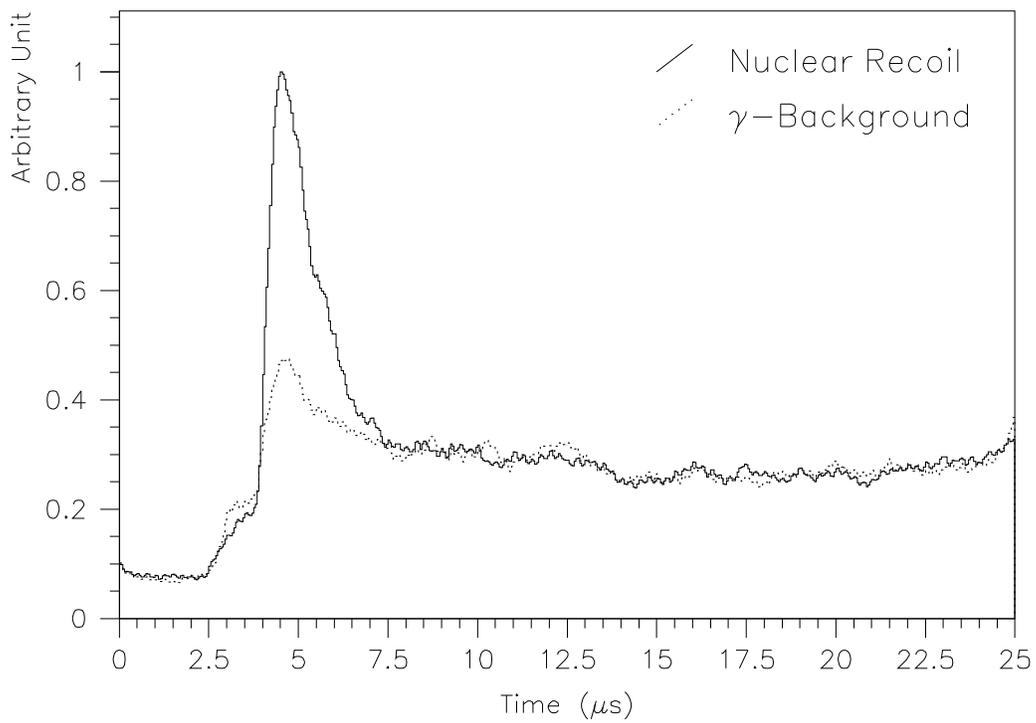,width=15cm}
}
\caption{
The average pulse shapes 
at 4.8~keV electron-equivalence energy for the
nuclear recoil and $\gamma$-background events 
directly from the neutron beam measurements.
}
\label{nbeam}
\end{figure}

The nuclear recoil pulses recorded in a neutron
beam environment were contaminated by
an intense accidental $\gamma$-background.
The average pulse shapes for both 
nuclear recoil and 
$\gamma$-background (as identified by
the ToF cut) events
derived from the neutron beam measurements
at a nuclear recoil e.e. energy of 4.8~keV
are depicted in Figure~\ref{nbeam}.
The long tails indicate there is
a substantial contribution
from time-uncorrelated $\gamma$-background.
Upon taking averages from a large sample 
and subtracting the $\gamma$-background, 
such data are sufficient to provide a 
good quenching factor measurement as well
as the average {\it ``background-free''} 
nuclear recoil pulse shapes, as
depicted in Figures~\ref{psdshape}a\&b.

\subsection{Average Light Emission Pulse Profiles}

From the FADC measurements with $\gamma$-events
and the subtraction procedures for recoil events
discussed in the previous section,
the {\it average} pulse shapes for both categories
are depicted in Figures~\ref{psdshape}a\&b.

In Figure~\ref{psdshape}a, the spread among
the average $\gamma$-pulses between 5~keV and 40~keV
are denoted by dotted lines, as compared 
to the solid line for the recoil pulse shape
at 4.8~keV e.e. energy (43~keV
recoil energy). Similarly, the spread among
recoil pulse shapes between 4~keV and 11~keV 
e.e. energy (about 40~keV to 110~keV recoil energy)
are defined by the dotted lines
in Figure~\ref{psdshape}b, whereas the
solid line denotes $\gamma$-profiles at 20~keV.
It can be concluded that
the energy dependence of the pulse shapes
{\it within} the recoil and $\gamma$-data samples
at this low ($<$100~keV) energy range
is small compared to the differences in the pulse
shapes {\it between} the recoil and $\gamma$-events.
These results are consistent with those
from Refs.~\cite{csidmfrance,csidmuk,csidmkorea}.
Event identification is in principle possible
be achieved at low energy with 
CsI(Tl) crystals 
from the pulse shape information.

The pulse shapes(A) as a function of time(t)
displayed in Figures~\ref{psdshape}a\&b
can be fitted
to an analytical form 
\begin{equation}
\rm{
A ~ =  Constant \ast [ ~ 1 - exp ( - \frac{t}{\tau_0} ) ~  ]
\ast
[ ~ \frac{1}{\tau_1} ~ exp ( - \frac{t}{\tau_1} )
+ \frac{r}{\tau_2} ~ exp ( - \frac{t}{\tau_2} ) ~ ] ~~,
}
\end{equation}
where  $\rm{\tau_0}$ is the rise time,
($\rm{\tau_1 , \tau_2}$) denote the fall times,
and r is the ratio between the slow and fast decay
components.
As illustrations of the typical ranges,
the best-fit values for recoil events at
4.8~keV e.e. energy
and $\gamma$-events at 20~keV
are tabulated in Table~\ref{time}.
Those for undoped CsI crystal are also shown for comparison.
The values of $\rm{\tau_0}$ in CsI(Tl) are dominated by
the electronics shaping rise time
of 250~ns for $> \mu$s pulses~\cite{electronics}.
The intrinsic rise times of the CsI(Tl)
scintillator are expected to $\sim$125~ns and $\sim$20~ns for
$\gamma$- and recoil-events, respectively~\cite{csichar}.
The difference in the decay time constants 
between the recoil and $\gamma$-events
is the basis of pulse shape discrimination.

\subsection{Data Samples}

The direct measurements of 
Figure~\ref{nbeam} indicate that
the {\it average} recoil pulse shape can be derived
by statistical subtraction.
However, at the event-by-event level,
the time-profile for photo-electron emissions
of the neutron beam data set
is complicated by an uncontrolled
and sizable background contribution.
Dark Matter
searches, on the other hand, 
are low-count-rate experiments
such that nuclear recoils due to WIMP interactions
will {\it not} be contaminated by accidentals.
Therefore the neutron beam data do not provide
a realistic sample for the studies of
detector response and realistic signals
in WIMP searches at
the event-by-event level.

\begin{figure}
\centerline{
\epsfig{file=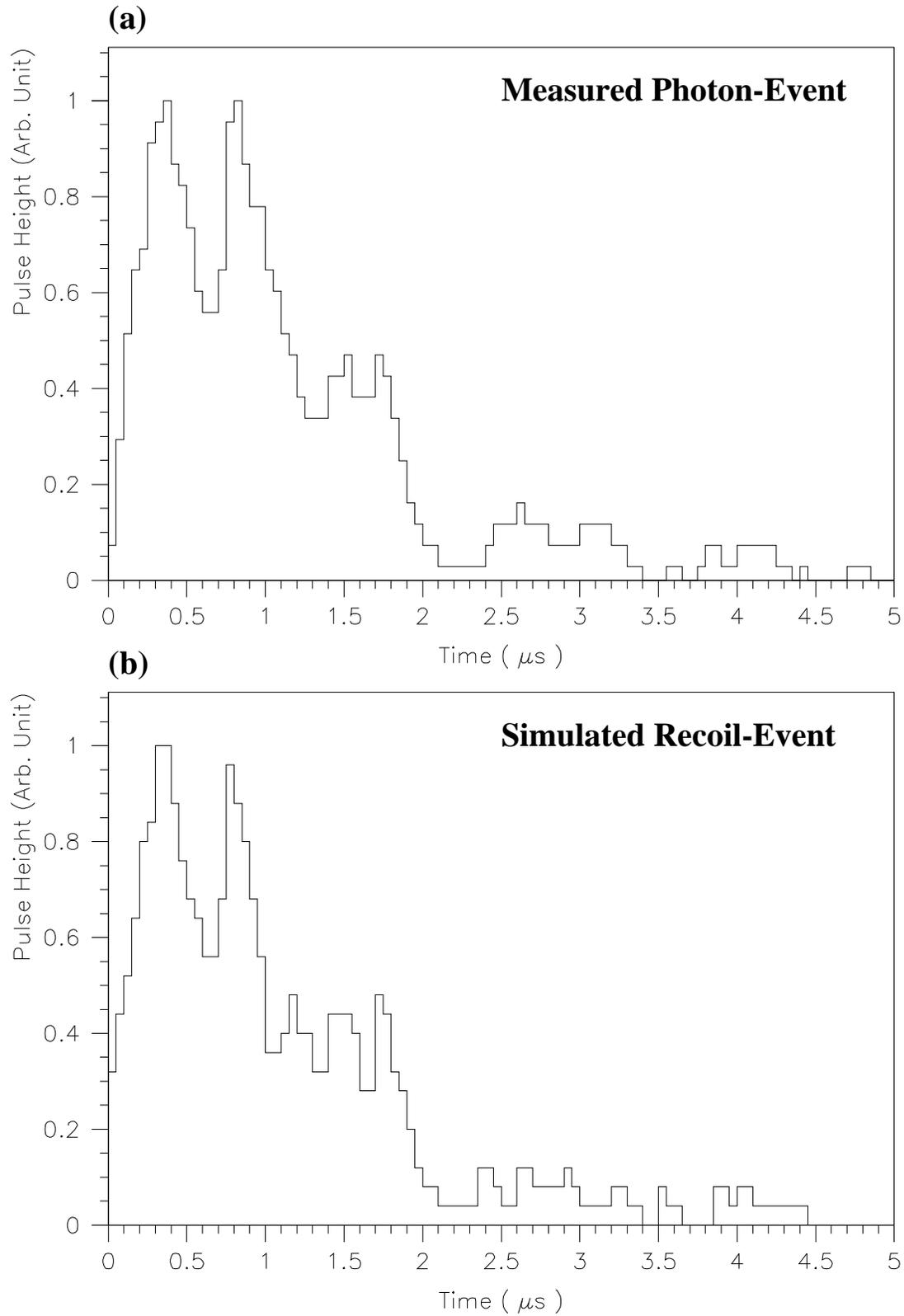,width=15cm}
}
\caption{
Typical single
(a) measured $\gamma$- and (b) simulated
nuclear recoil events at $\npe \sim 20$. 
}
\label{singlepulse}
\end{figure}

As remedies, the single-event
nuclear recoil pulse shape
was generated by simulations,
where the input included
the measured pulse profile
at 4.8~keV e.e. energy from Figure~\ref{psdshape}a
and the parametrization of Table~\ref{time}.
The simulated events were 
convolutions of (a) a total of $\npe$
single photo-electron pulses whose timing
fluctuations were generated according to the 
average recoil pulse shape, and
(b) the single photo-electron response of
the PMT and readout system for each of these pulses, 
provided by the LED pulser measurements.
A ``self-trigger'' criterion was imposed
to mimic the realistic situation
$-$ that is,
the time-zero of the events was defined
by the first instants where the pulse was above
a specified threshold.
As illustrations,
typical events at $\npe \sim 20$
from the measured $\gamma$ and simulated 
nuclear recoil data
are displayed in Figures~\ref{singlepulse}a\&b,
respectively.  Both categories
of events are similar by visual inspection, 
demonstrating that
(a) the simulation algorithms are valid, and
(b) advanced pattern recognition techniques
would be necessary to achieve event identification.

\begin{figure}
{\bf (a)}\\
\centerline{
\epsfig{file=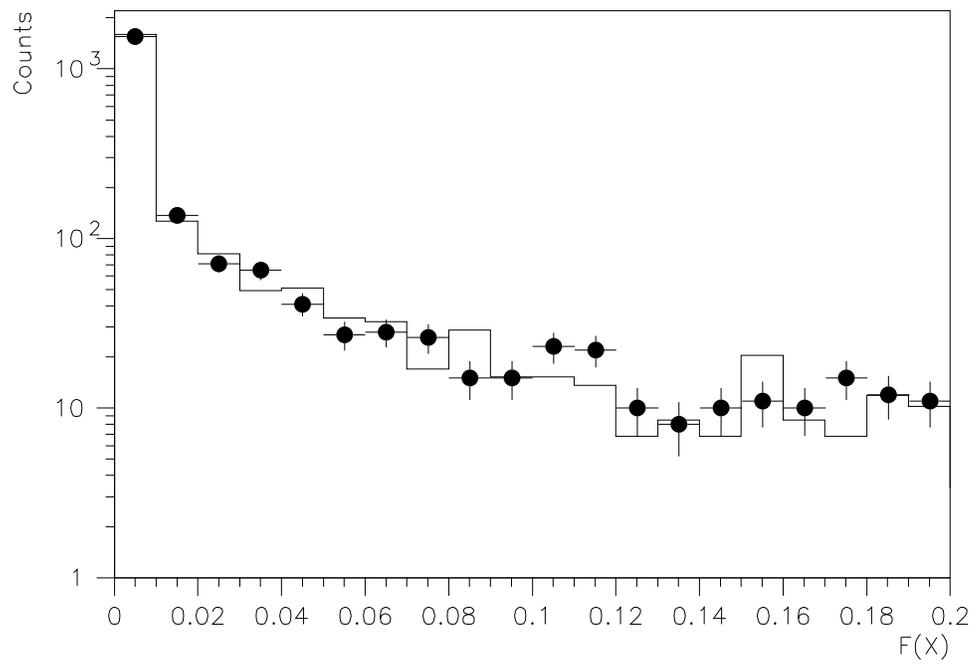,width=14cm}
}
{\bf (b)}\\
\centerline{
\epsfig{file=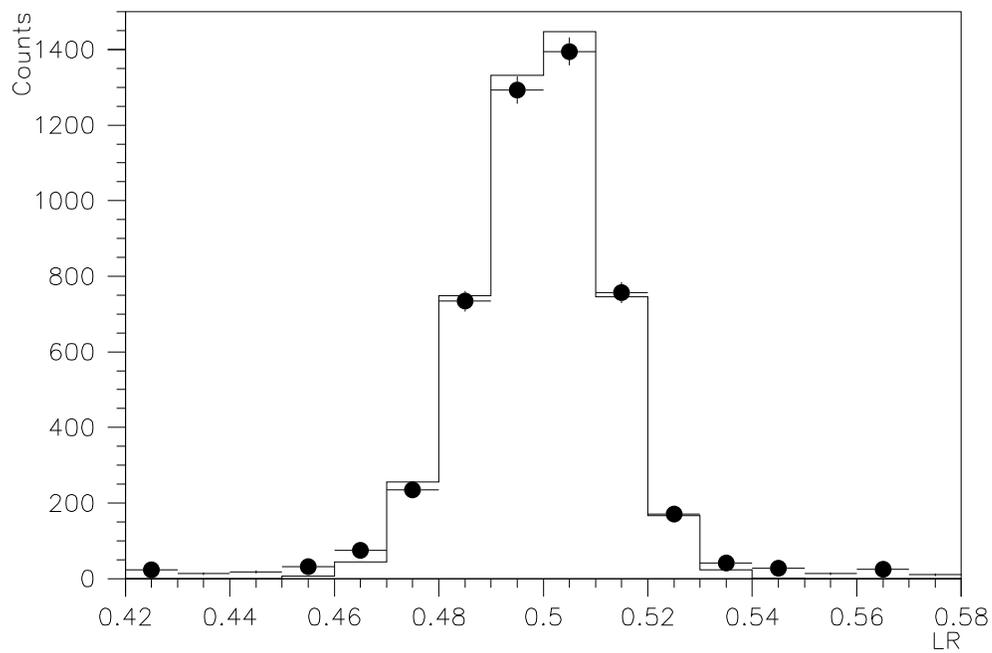,width=14cm}
}
\caption{
Comparison of the distributions
of (a) F(X) and (b) LR parameters
between measured and simulated 
data denoted by solid circles and
histograms, respectively.
}
\label{compare}
\end{figure}

Applying the same algorithm on the
$\gamma$-reference profile at 20~keV 
shown in Figure~\ref{psdshape}b,
simulated $\gamma$-events were also generated. 
The comparisons between the distributions of the PSD
figures-of-merit among the simulated and
measured $\gamma$-events at $\npe \sim 60$
are depicted in Figures~\ref{compare}a\&b. 
The agreement is excellent,
further demonstrating the ability
of the pulse shape simulation algorithms
in reproducing correctly
the fluctuations among individual events. 
The figures-of-merit will be defined and
discussed in details in Section~\ref{sect::psd}.

The excellent consistencies in 
Figures~\ref{compare}a\&b
between simulations and data for the 
$\npe \sim 60$ $\gamma$-events
justifies the use of the
simulated nuclear recoil and $\gamma$ data set
(denoted by $\lenr$ and $\leg$, respectively)
for the PSD studies discussed in subsequent sections. 
There are better uniformity and systematic control
among the different simulated data set such that
the residual systematic
effects will be  minimized
and canceled out when comparisons are made.
The input parameters to the simulation procedures
can be varied to study features like
sensitivities and robustness.

\subsection{Classical Pulse Shape Discrimination Method}
\label{sect::psd}

\subsubsection{Double Charge Method}

A well-established way to achieve PSD 
at high light yield is 
the ``double charge method''~\cite{psddc}.
This involves the comparison of 
the ``total charge'' ($\rm{Q_t}$) 
and the ``partial charge'' ($\rm{Q_p}$),
which are the total and partial
integration of the pulse, respectively.
This is the standard approach with
Analog Digital Convertor (ADC) based
data acquisition systems where the
complete pulse shape information
is not available. Typically,
the partial charge measurement is done
by delaying the PMT pulses via cabling 
and both the prompt and delayed
signals are read out by the ABC
sampled with the same gate.

\begin{figure}
\centerline{
\epsfig{file=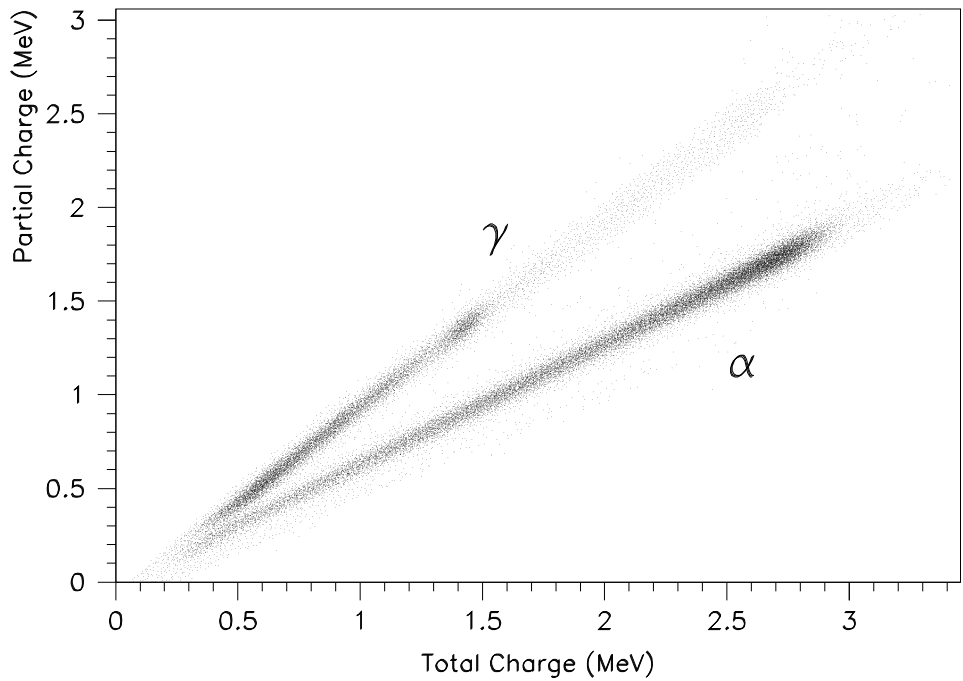,width=15cm}
}
\caption{
The partial charge versus total charge
at the high (MeV) energy range 
in a CsI(Tl) crystal,
showing excellent ($>$99\%) pulse shape discrimination
capabilities to differentiate events due to
$\alpha$'s and $\gamma$'s.
The $\alpha$-events are from
an $^{241}$Am source (kinetic energy 5.49~MeV)
placed on the surface of the
crystal, while the $\gamma$-events are due
to ambient radioactivity. 
}
\label{psd2dhe}
\end{figure}

Displayed in Figure~\ref{psd2dhe}
is the comparison of $\gamma$ and $\alpha$  
events at the MeV energy range
from data with ambient radioactivity
and $^{241}$Am $\alpha$-source,
respectively.
The ranges were chosen such that
$\rm{Q_t}$ and $\rm{Q_p}$ involve integration
over 4~$\mu$s after trigger and
{\it after} a delay of 0.5~$\mu$s, respectively.
A $\gamma$/$\alpha$ separation 
of $>$99\% efficiency down to
about 200~keV e.e. light output
can be achieved.
It has been shown that PSD can 
be achieved even in high energy events
where the FADC measurements are 
saturated~\cite{drange}.
However, as indicated in Figure~\ref{psd2dhe},
one would come into difficulties to perform
PSD with this simple algorithm 
for events at light yield 
below 100~keV e.e.
energy.

\subsection{Pulse Shape Discrimination Methods at Full Digitization} 

With the advent and popular usage of FADCs, 
complete pulse shape digitization becomes realistic.
Three different pattern recognition
techniques were investigated,
all of which rely on the
full digitization of the PMT signals.

\subsubsection{Mean Time Method}

The measurement of the average time for
individual events by the
mean time (MT) method has been used
for PSD studies~\cite{csidmfrance}.
The mean time is defined as 
\begin{equation}
\rm{
\langle t \rangle ~ = ~
 { \sum\limits_{i} ~ ( A_i ~  t_i ) \over \sum\limits_{i} ~ A_i }  ~~ ,
}
\end{equation}
where $\rm{A _i}$ is the FADC-amplitude at
time-bin $\rm{t _i}$. 

\begin{figure}
{\bf (a)}\\
\centerline{
\epsfig{file=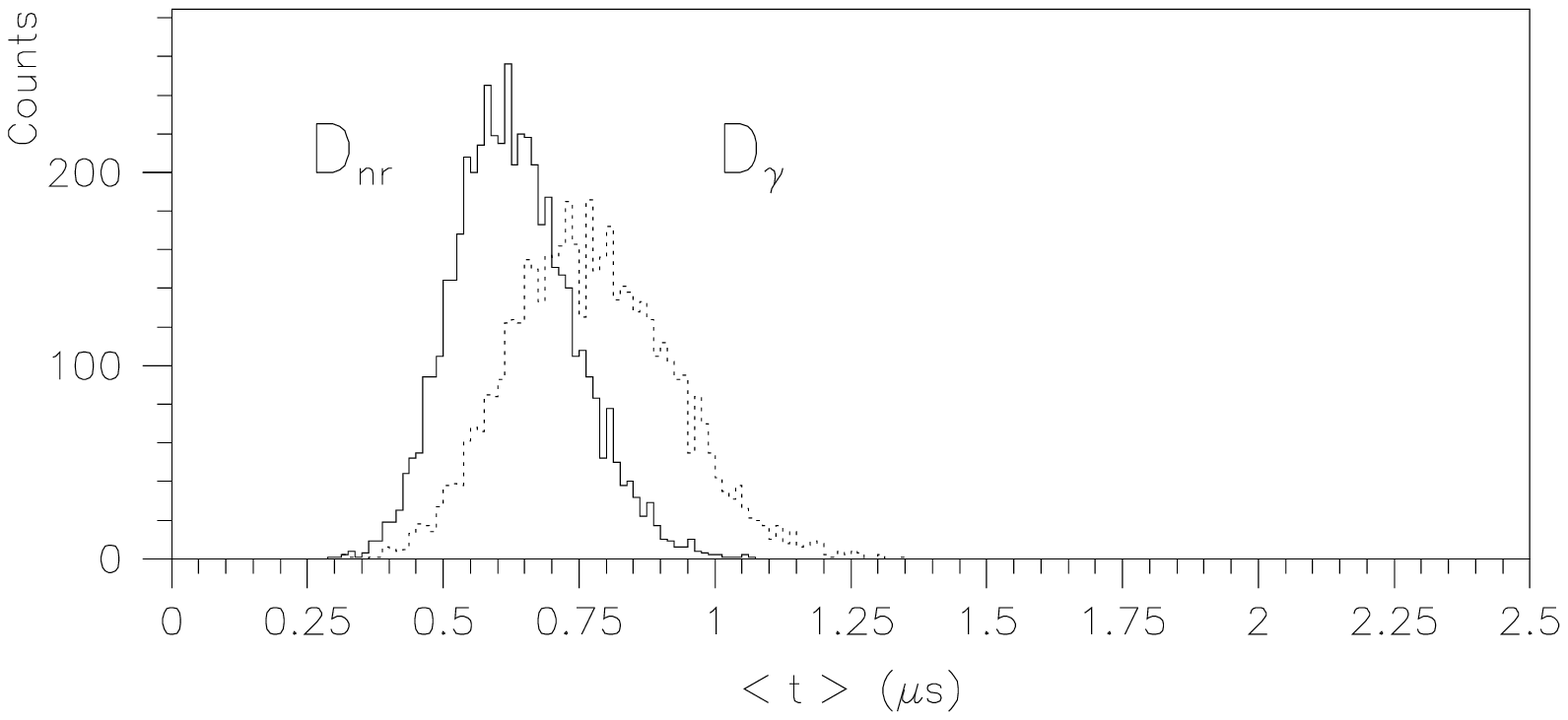,width=12cm}
}
{\bf (b)}\\
\centerline{
\epsfig{file=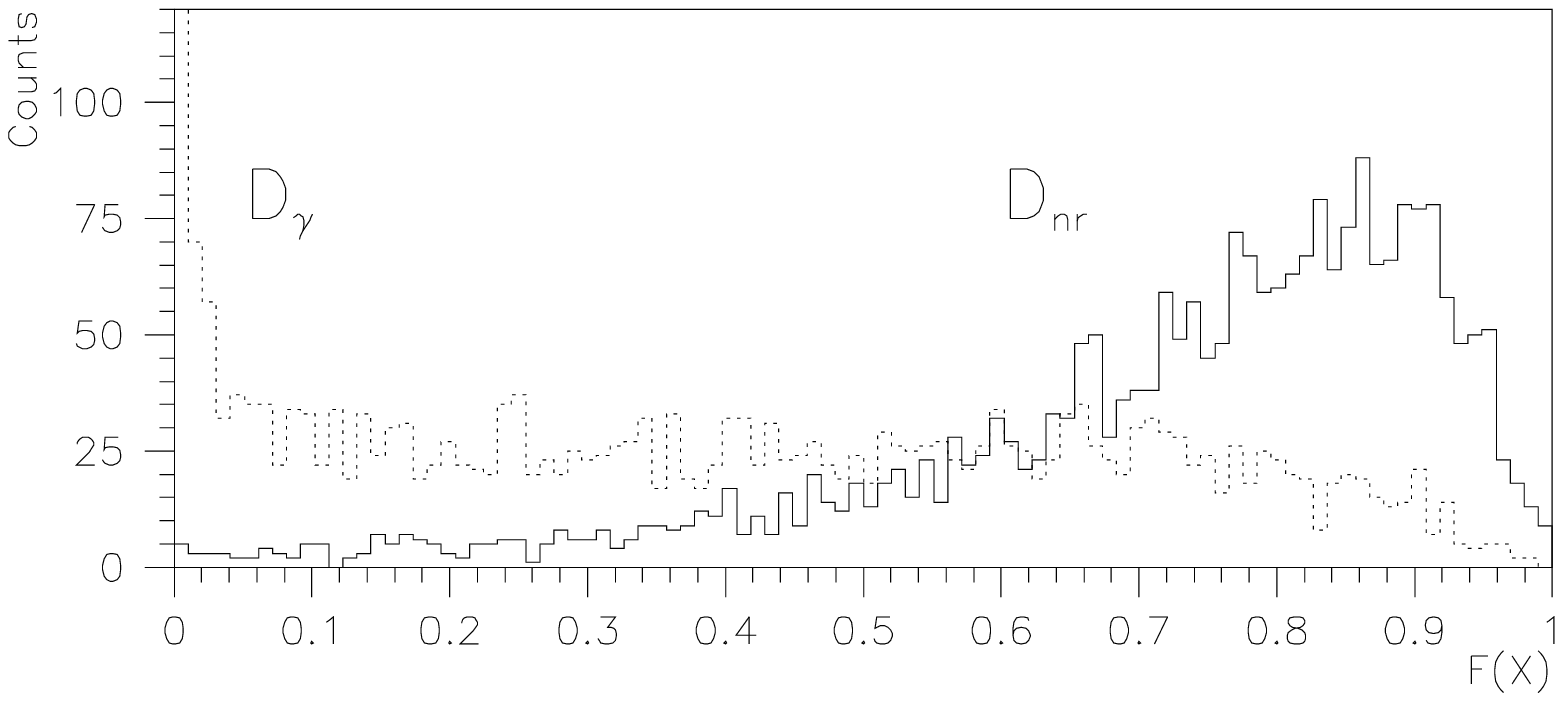,width=12cm}
}
{\bf (c)}\\
\centerline{
\epsfig{file=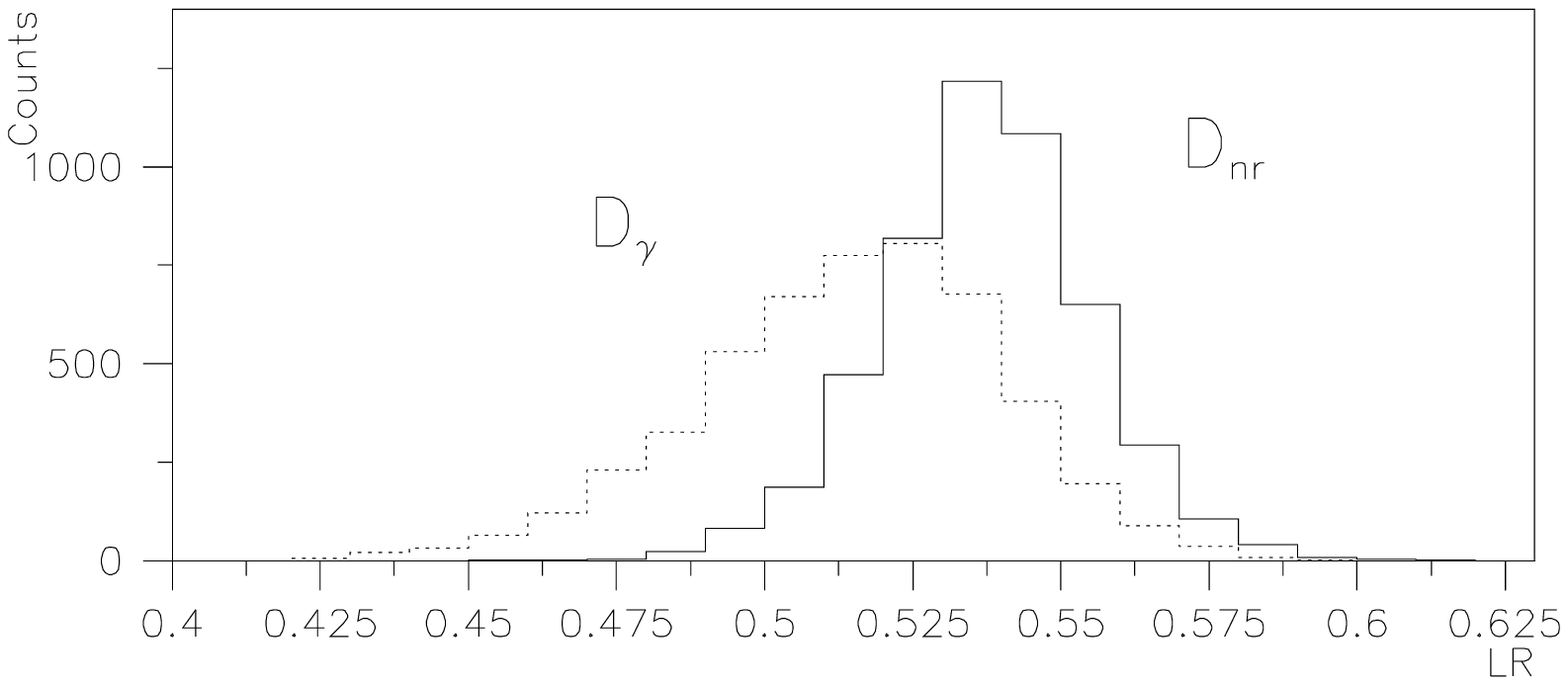,width=12cm}
}
\caption{
Typical separations of the 
(a) $\rm{\langle t \rangle}$,
(b) $\rm{ F ( X )}$ ,
and (c) LR parameters
at $\npe = 20$
between nuclear recoil ($\lenr$, in solid histograms)
and $\gamma$ ($\leg$, in dotted histograms)
events with the mean time, 
neural network 
and likelihood ratio methods,
respectively.
}
\label{single}
\end{figure}

The typical $\rm{\langle t \rangle}$ distributions
at $\npe \sim 20$ for $\lenr$ and $\leg$
are displayed in Figure~\ref{single}a,
at an integration of 5~$\mu$s after the
time-zero set by the trigger.
It can be seen that satisfactory separation
can be achieved under such conditions.

\subsubsection{Neural Network Methods}
\label{sect::neuralnet}

The neural network (NN) methods~\cite{neuralnet} 
are now frequently
adopted for analysis in high energy physics experiments.
It has been applied to event-by-event pulse
shape analysis for background identification in 
double beta decay searches~\cite{dbd}.
The pedestal-subtracted FADC 
data within 5~$\mu$s after trigger
corresponds to the input nodes
of the neural network.
That is, the network has
$\nin$=100 input nodes  denoted by $\rm{X ( x_i )}$
with the integrated sum 
normalized to unity:
\begin{equation}
\rm{ \sum_{i=1}^{\nin} x_i = 1 }  ~ .
\end{equation}
Negative values were reset to zero.
In addition, the number of hidden nodes was selected
to be $\nhid$=25. It has been checked that the
results are independent of this choice, so long
as $\nhid > 20$.

Adopting the Neural Network 
JETNET 3.0 package~\cite{neuralnet},
a function F(X) is defined such that
\begin{equation}
\rm{
F(X) ~ = ~
G (  \sum_{j=1}^{ \nhid } u_j  ~
G (  \sum_{k=1}^{ \nin }  w_{jk}  ~ x_k + \theta _j )
+ \phi _0 )
}
\end{equation}
where 
$\rm{(  u_j , ~ w_{jk} )}$
and $\rm{( \theta _j ,  \phi _0 )}$ are the
``weight'' and 
``offset'' coefficients, respectively,
to be derived from the
training samples, and the function G(y)
is the non-linear neuron activation function 
\begin{equation}
\rm{
G(y) ~ = ~
\frac{1}{2} ~ [ ~ 1 + tanh (y) ~ ] ~ = ~ 
\frac{1}{ 1 + e^{-2y} } ~~~~~ ,
}
\end{equation}
which is the functional form characterizing
a 3-layer neural network consisting of 
the input, hidden and output layers.

A total of $\ntr$=4000  events
from both the $\lenr$ and $\leg$ data sets
are used as training samples,
corresponding to T($X$)=1 and 0, respectively.
The optimal coefficients 
are obtained by minimizing
the error function
\begin{equation}
\rm{
E ~ = ~
\sum_{i=1}^{ \ntr } ~ [ ~ F ( X ) - T ( X ) ~ ] ^2  ~~~ .
}
\end{equation}

Once the coefficients are derived, the procedures
are applied to {\it independent} recoil and $\gamma$
data set.
The F(X)-values for recoil events
would be larger than those for $\gamma$-events
at the same light yield.
The comparisons of F(X) distributions
for the simulated and
measured $\gamma$-data at $\npe \sim 60$
are shown in Figure~\ref{compare}a,
while  those  
for $\lenr$ and $\leg$ at $\npe \sim 20$
are displayed in Figure~\ref{single}b.
It can be seen that the simulations 
agree well with data, while there is
good separation between the recoil
and $\gamma$ samples.

\subsubsection{Likelihood Ratio Methods}

Motivated by the commonly-used of
likelihood ratio test~\cite{pdg,lratio} for the 
goodness-of-fit, 
a likelihood ratio (LR) method was devised 
to perform the tasks of pulse shape analysis.
Similar methods are successfully
applied in high
energy physics data analysis in
comparing likelihoods and 
assigning probabilities 
among the different hypotheses 
for events where many output parameters 
are measured.
The reference profiles 
for neutrons and $\gamma$'s 
from Figure~\ref{psdshape} are 
required as the input.
This is different 
from the previous two techniques 
where prior knowledge of the
reference profiles is not necessary.

The areas of the reference pulses
are normalized to unity, and the
profiles are denoted by arrays
$\rm{R ( r _i) }$ and
$\rm{\Gamma (\gamma _i) }$ for
the nuclear recoil and $\gamma$ reference shapes,
respectively.
Two likelihood functions, $\rm{L_{r}}$ and $\rm{L_{\gamma}}$,
are defined for each event:
\begin{equation}
\rm{
L_{r} ~ = ~ \prod_{i=1}^{\nin} ~ r _i ^{x_i}  ~~~~ ; ~~~~~
L_{\gamma} ~ = ~ \prod_{i=1}^{\nin} ~ \gamma _i ^{x_i} ~~ ,
}
\end{equation}
where $\rm{X (x_i)}$ with dimension $\nin = 100$
are the measured pulse shape information 
for the events to be analyzed,
as defined in Section~\ref{sect::neuralnet}.
The likelihood functions quantify how
probable the measured pulse shapes 
do originate from the the reference profiles.
The likelihood ratio $\lr$ defined by:
\begin{equation}
\rm{
\lr ~ = ~ \frac{ L_{r} }{ L_{r} + L_{\gamma} } ~~,
}
\end{equation}
will test which hypothesis is more likely.
The LR-values for recoil events
would be larger than those for $\gamma$-events
at the same light yield.

The comparisons of $\lr$ between simulated and
measured $\gamma$-data at $\npe \sim 60$
are shown in Figure~\ref{compare}b.
while the typical $\lr$ distributions
for the $\lenr$ and $\leg$ data set
at $\npe \sim 20$
are depicted in Figure~\ref{single}c. 
Similar to the neural network methods,
there are good agreement between
simulations and data,
while there is satisfactory
separation between the recoil
and $\gamma$ samples.

\subsection{Comparisons}

The excellent agreement depicted in 
Figures~\ref{compare}a\&b 
between measured and simulated $\gamma$-data set
justifies that valid comparisons 
can be made between
the simulated data set $\leg$ and $\lenr$.
To quantify, two figures of merits are defined:
(a) $\e90$: the survival efficiencies of
$\lenr$ at selections which
ensure that 90\% of the $\leg$ events
are suppressed; and
(b) $\l90$: the probabilities where the
$\leg$ events would be mis-identified
as recoil signals at cuts where
90\% of $\lenr$ would survive.
Both $\e90$ and $\l90$ are energy dependent,
and would approach 1 and 0, respectively,
at the high light yield (large $\npe$) limits.

The variations of $\e90$ and $\l90$ as
a function of $\npe$ for the three different
methods (MT, NN, LR) are depicted
in Figures~\ref{fom}a\&b,
respectively.
The photo-electron number $\npe$ was adopted
as the unit to characterize the light yield.
In this way,
the results can be directly
applicable to other configurations
using CsI(Tl) as the detector medium.
Dotted lines in Figures~\ref{fom}a\&b
corresponds to the survival probabilities of
$\leg$ and $\lenr$, respectively.
The results indicate that the
generic features that all the
three methods:
(a) can achieve PSD with satisfactory
efficiencies ($>$50\% $\gamma$-background rejection) 
at $\npe > 20$;
(b) can identify $>$90\% of the $\leg$ background
while keeping  the efficiencies for $\lenr$
to be $>$90\% at $\npe >  80$;  and
(c) give similar performance at the
large light yield limit ($\npe = 120$),
which approaches the expected values
of 1 and 0 for $\e90$ and $\l90$, respectively.
Comparing the relative merits among 
the three algorithms,
the NN technique gives better performance 
in the low energy range.
The relative merits between the
LR and MT methods are similar at
low statistics ($\npe < 20$), while LR tends
to perform better at the intermediate range
($40 < \npe < 80$). 
Accordingly, the neural network method
would be the preferred technique in
pulse shape analysis at regime where
the statistics is marginal.

Tests have been performed on simulated
events with different single photo-electron response
functions. Results consistent with the performance
parameters shown in Figures~\ref{fom}a\&b
were obtained. This shows that the results
are robust and insensitive to the details of
the simulation algorithms so long as the 
same reference profiles in Figures~\ref{psdshape}a\&b
are used for the photo-electron
timing distributions. 
This indicates that the results
would also be valid in measurements of CsI(Tl) crystals
where the  PMT response 
and electronics settings (such as 
shaping times) would be different.

\section{Summary and Conclusions}

This article reports the 
measurement of light emission pulse profiles
for nuclear recoil and $\gamma$-events in
CsI(Tl) crystal scintillator
at the energy range relevant for Dark Matter
searches ($<$ 100 keV).
The energy dependence of the profiles
within the recoil and $\gamma$-samples is small 
compared to the differences between them.
Event identification is feasible for CsI(Tl)
based on pulse shape discrimination.

Various software techniques
to achieve pulse shape 
discrimination in this 
``near threshold'' regime
were studied.
The performance of the three methods
based on complete pulse shape information
(mean time, neural network and likelihood ratio)
is superior to the matured and conventional
double charge method well-demonstrated when
photo-electrons are abundant.
Full digitization is crucial for 
achieving PSD at the marginal statistics
domain. 
Among the three methods studied,
the neural network technique provides the
best performance in regime where the
statistics is marginal, while
the other two methods are still 
satisfactory. 
The algorithms are robust and insensitive
to the measurement parameters like PMT response
or electronic shaping times.

The results from this study
are relevant to the
potential capabilities and
practical design of Dark Matter experiments
based on the CsI(Tl) crystal.
Satisfactory (bigger then 50\%) separation 
between $\gamma$ and
nuclear recoil events can
be achieved when the photo-electron
statistics is larger than 20,
which corresponds to an 
electron-equivalence energy threshold of about 5~keV,
or 40$-$50~keV recoil kinetic energy,
in the adopted detector configuration
of 0.56~kg target mass.
In realistic Dark Matter experiments,
the modular mass for the CsI(Tl) target
will have to be bigger, such as 
at the range of several kg.
To maintain or even improve on such
threshold, the light transmission
within the crystal and the optical
coupling between the crystal surface
and the PMT photo-cathode 
will have to be optimized.
Larger PMT readout surfaces 
as well as green-extended
photo-cathodes to match the spectral
emissions of CsI(Tl) can be used.
Photon detectors with higher quantum
efficiencies, such as avalanche photo-diodes, 
can also be considered, though these devices
tend to be limited in the active surface area.

\begin{figure}
\centerline{
\epsfig{file=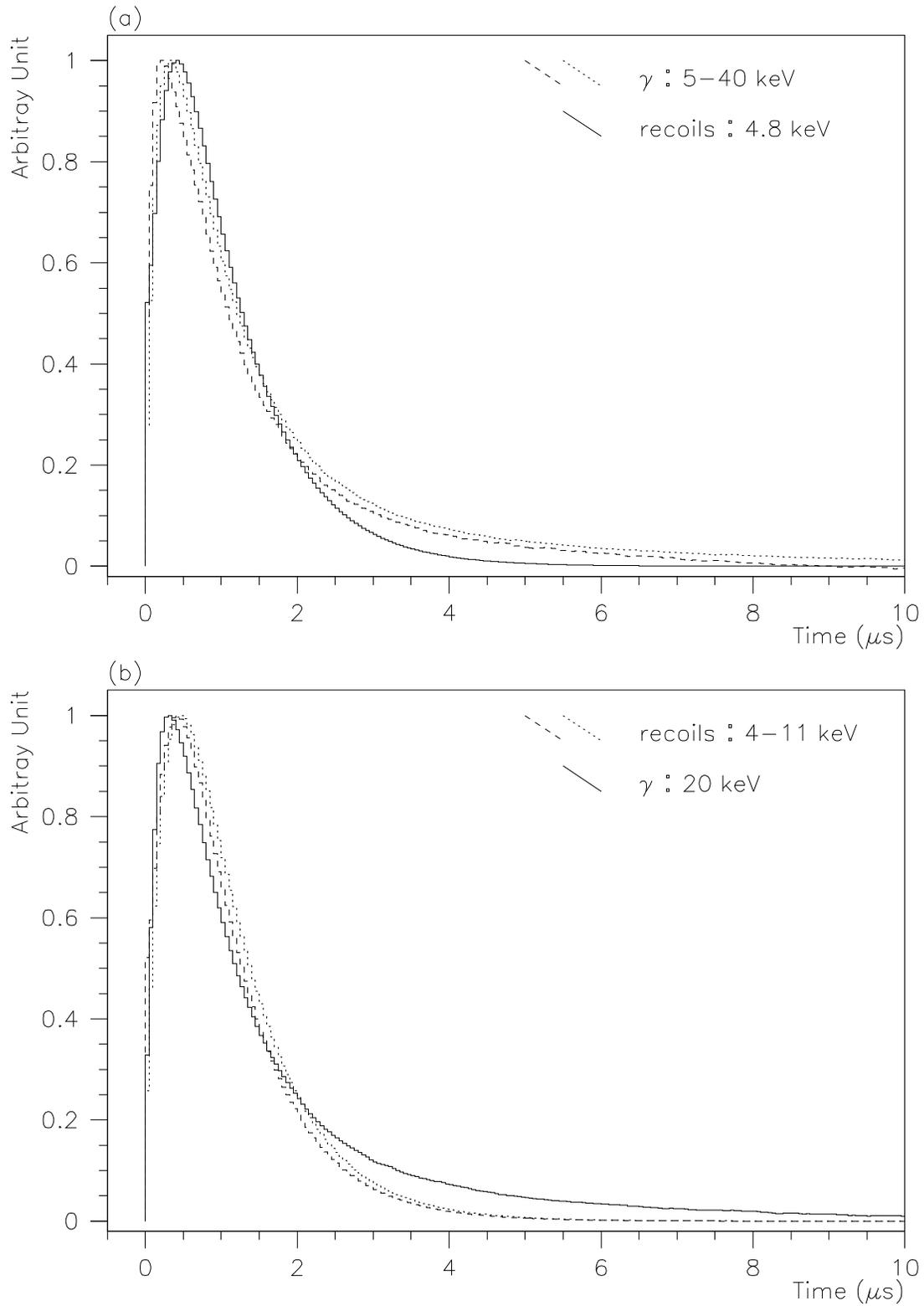,width=15cm}
}
\caption{
Comparison of average pulse shapes
between $\gamma$- and
nuclear recoil events:  (a) recoil pulse
at 4.8~keV electron-equivalence energy
as compared to $\gamma$-events from 5~keV
to 40~keV; (b) $\gamma$-pulse at 20~keV
as compared to recoil profiles from
4~keV to 11~keV electron-equivalence
energy.
}
\label{psdshape}
\end{figure}

\begin{figure}
{\bf (a)}\\
\centerline{
\epsfig{file=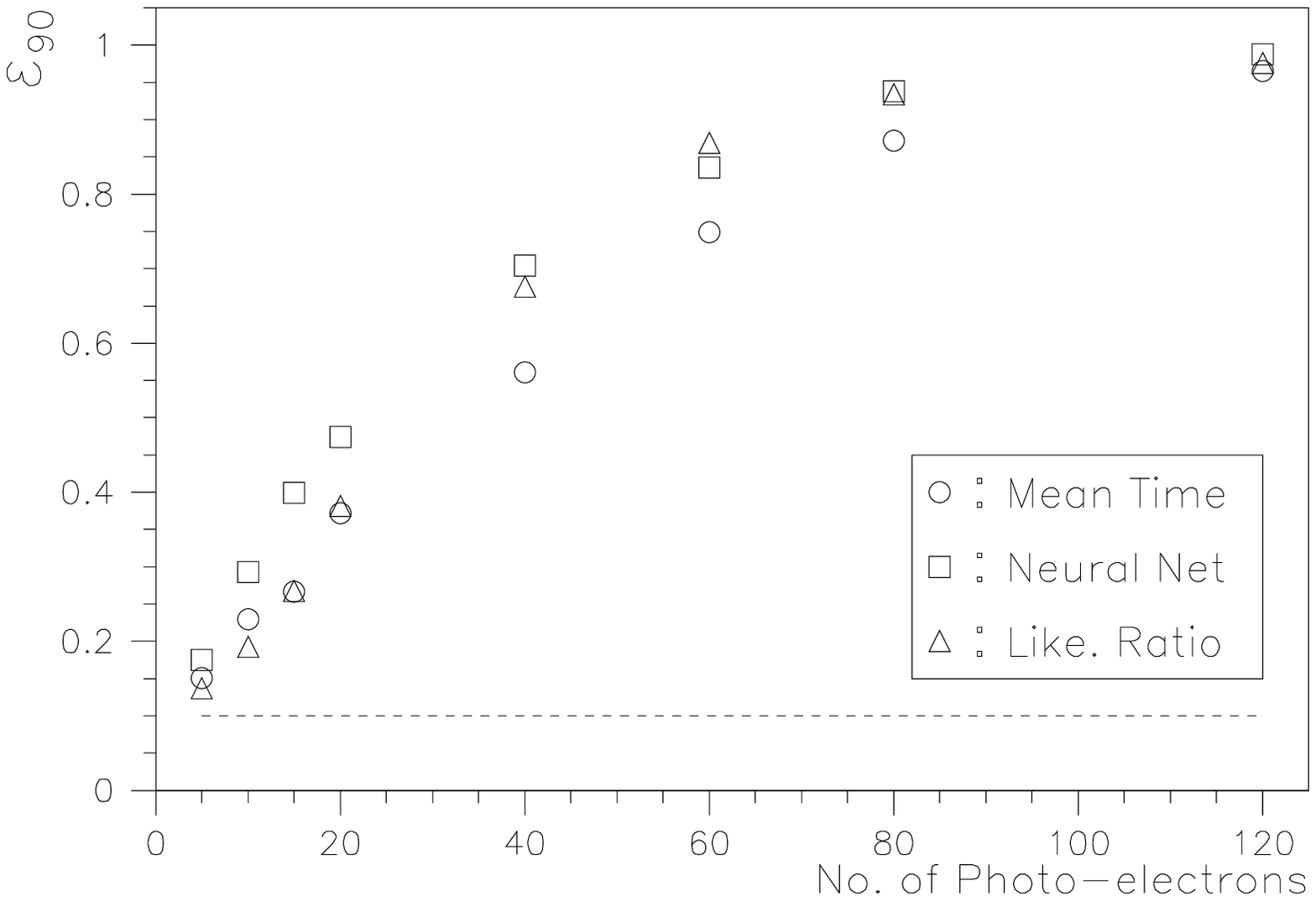,width=14cm}
}
{\bf (b)}\\
\centerline{
\epsfig{file=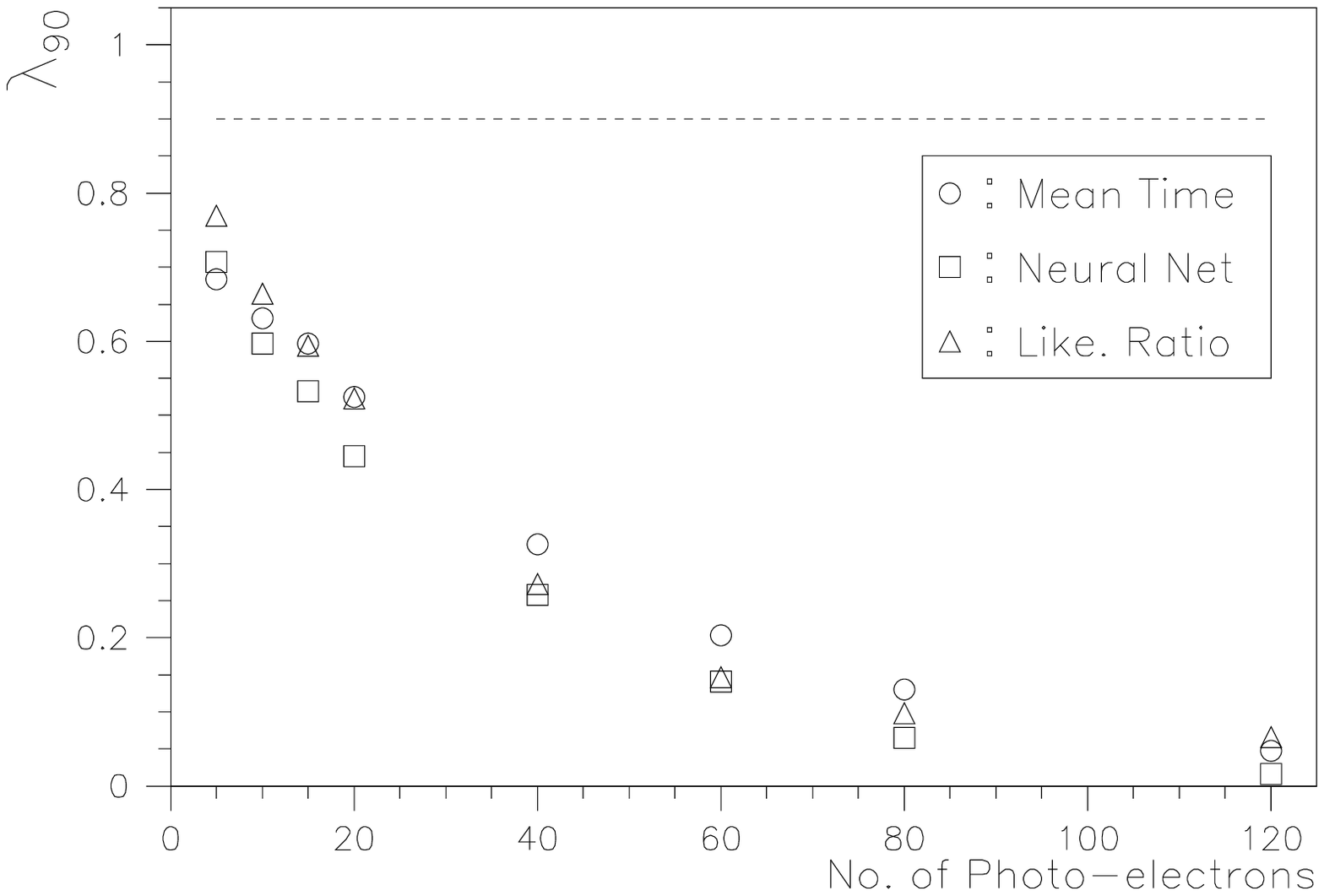,width=14cm}
}
\caption{
The variations of the figures of merit (a) $\e90$ 
and (b) $\l90$ with $\npe$ with the
three different techniques applied to simulated
nuclear recoil and $\gamma$ data, respectively. 
Dotted lines indicate survival probabilities of
$\gamma$ and recoil events in (a) and (b), respectively.
The statistical uncertainties are smaller than the data points.
}
\label{fom}
\end{figure}

Although this article focuses on data with
CsI(Tl) crystal scintillators for Dark
Matter searches, the software
techniques are readily applicable to other
detector systems for other experiments
where pulse shape analysis of individual 
events can provide useful information.
A universal reference pulse
profile was adopted in the present studies,
one for recoil and one for
$\gamma$ events, for the entire energy range
of interest. This is justified in view of
Figures~\ref{psdshape}a\&b such that 
the results in
Figures~\ref{fom}a\&b are 
valid for CsI(Tl) crystals. 
In other detector systems where there
may be a stronger energy dependence of the pulse shapes
(as in the case for $\gamma$-events
with NaI(Tl) crystals~\cite{csidmfrance}),
the PSD procedures can be further refined by
having a different reference profile for every energy
bin. The profiles can be derived from measurements
with neutron beam and radioactive $\gamma$-sources
like in this work.
In addition, the conclusions on the
relative merits among the different PSD methods
are expected to be applicable to
other pulse shape analysis problems where
the statistics are marginal.

Besides differentiating
$\beta$/$\gamma$-background from nuclear recoil events,
these studies may help 
to lower the detection threshold
by suppressing electronic noise and
microphonism where the pulse shapes
are in general
different from those of the signals.
Experiments which need {\it both}
low threshold and background 
may potentially benefit from these techniques.
Alongside with Dark Matter experiments,
such requirements are critical in the 
search of neutrino magnetic moments~\cite{ksexpt}
and in the measurement of 
the coherent scatterings
of the neutrinos on the nuclei~\cite{cohsc}.

The authors would like to thank
Drs. S.K. Kim and Y.D. Kim
for fruitful discussions and helpful comments,
and are grateful to
the technical staff
from CIAE and IHEP for assistance in the
neutron beam data taking.
This work was supported by contracts
CosPa~89-N-FA01-1-4-2 from the Ministry of Education, Taiwan,
NSC~89-2112-M-001-056, 
NSC~90-2112-M-001-037
and 
NSC~91-2112-M-001-036
from the National Science Council, Taiwan,
and NSF19975050 from the
National Science Foundation, China.

\clearpage

\begin{table}
\begin{center}
\begin{tabular}{|c||c|c|c|c|c|}
\hline
& & & \multicolumn{2}{|c|}{ } & \\
Crystal & Event Type & Rise Time & 
\multicolumn{2}{|c|}{Decay Time Constant}
& Ratio (r) \\ \cline{4-5}
&  & [$\rm{\tau_0}$ (ns)] & Fast Comp. 
& Slow Comp. & \\ 
& & & 
[$\rm{\tau_1}$ ($\mu$s)] & [$\rm{\tau_2}$ ($\mu$s)] & \\ \hline \hline
& & & & & \\
CsI(Tl) & nuclear recoils & 280$\pm$50 & 0.54$\pm$0.1  & 
2.0$\pm$0.2  & 0.29$\pm$0.02  \\
& & & & & \\
CsI(Tl) & $\gamma$ & 296$\pm$50 & 1.3$\pm$0.1  & 
4.50$\pm$0.4  & 0.58$\pm$0.06 \\ 
& & & & & \\
CsI(pure) & $\gamma$ & $\sim$0.55  & 0.19$\pm$0.02  & $-$  & $-$ \\ 
& & & & & \\ \hline
\end{tabular}
\end{center}
\caption{
Fitted rise and decay time constants as
well as the ratio between slow and fast decay components
for recoil and $\gamma$ events measured by CsI(Tl)
and undoped CsI.
}
\label{time}
\end{table}

\clearpage

\clearpage

\clearpage

\clearpage

\clearpage

\clearpage 

\clearpage

\clearpage

\begin{thebibliography}{99}

\bibitem{pdg}
See the respective sections in
{\it Review of Particle Physics},
Particle Data Group, Phys. Rev. {\bf D 66} (2002),
for details and references.
\bibitem{csichar}
H. Grassmann, E. Lorentz and H.G. Moser,
Nucl. Instrum. Methods {\bf 228}, 323 (1985);\\
P. Schotanus, R. Kamermans, and P. Dorenbos,
IEEE Trans. Nucl. Sci. {\bf 37}, 177 (1990).
\bibitem{prospects}
H.T. Wong et al.,
Astropart. Phys. {\bf 14}, 141 (2000).
\bibitem{ksexpt}
H.T.~Wong and J. Li,
Mod. Phys. Lett. {\bf A 15}, 2011 (2000);\\
H.B. Li et al., TEXONO Coll.,
Nucl. Instrum. Methods {\bf A 459}, 93 (2001);\\
H.B. Li et al., TEXONO Coll.,
Phys. Rev. Lett. {\bf 90}, 131802 (2003).
\bibitem{proto} 
Y. Liu et al., TEXONO Coll., 
Nucl. Instrum. Methods {\bf A 482}, 125 (2002).
\bibitem{csidmfrance} 
G.~Gerbier et al., Astropart. Phys. {\bf 11}, 287 (1999);\\
S.~Pecourt et al., Astropart. Phys. {\bf 11}, 457 (1999).
\bibitem{csidmuk} 
V.A. Kudryavtsev et al., Nucl. Instrum. Methods {\bf A 456}, 272 (2001).
\bibitem{qfpaper}
M.Z.~Wang et al., Phys. Lett. {\bf B 536}, 203 (2002).
\bibitem{csidmkorea} 
H. Park et al., Nucl. Instrum. Methods {\bf A 491}, 460 (2002);\\
T.Y. Kim et al., Nucl. Instrum. Methods {\bf A 500}, 337 (2003).
\bibitem{dama}
R. Bernabei et al., Phys. Lett.{\bf B 480}, 23 (2000),
and references therein.
\bibitem{bfactories}
Y. Kubota et al., CLEO Coll.,
Nucl. Instrum. Methods {\bf A 320}, 66 (1992);\\
E. Aker et al., Crystal Barrel Coll.,
Nucl. Instrum. Methods {\bf A 321}, 69 (1992);\\
K. Miyabayashi, Belle Coll.,
Nucl. Instrum. Methods {\bf A 494}, 298 (2002);\\
B. Lewandowski,  BaBar Coll.,
Nucl. Instrum. Methods {\bf A 494}, 303 (2002).
\bibitem{scinbasic}
See, for example,
J.B.~Birks,
{\it Theory and Practice of Scintillation
Counting}, Pergamon (1964).
\bibitem{psdpid}
J. Alarja et al., Nucl. Instrum. Methods {\bf A 242},
352 (1982);\\
F. Benrachi et al., Nucl. Instrum. Methods {\bf A 281},
137 (1989).
\bibitem{electronics}
W.P.~Lai et al., TEXONO Coll.,
Nucl. Instrum. Methods {\bf A 465}, 550 (2002).
\bibitem{psddc}
C.L. Morris et. al., Nucl. Instrum. Methods {\bf 137},
397 (1976);\\
M.S. Zucker and N. Tsoupas,
Nucl. Instrum. Methods {\bf A 299}, 281 (1990).
\bibitem{drange}
Q.~Yue et al.,  
Nucl. Instrum. Methods {\bf A 511}, 408 (2003).
\bibitem{neuralnet} 
See, for example,\\
C. Peterson, T. Rognvaldsson and L. Lonnblad, 
Comput. Phys. Comm. {\bf 81}, 185 (1994); \\
C.M.~Bishop, {\it Neural Networks for Pattern Recognition},
Clarendon Press, Oxford (1995).
\bibitem{dbd}
B. Majorovitis and H.V. Klapdor-Kleingrothaus,
Eur. Phys. J. {\bf A 6}, 463 (1999).
\bibitem{lratio}
S. Baker and R.D. Cousins, 
Nucl. Instrum. Methods {\bf 221}, 125 (1984), and references
therein.
\bibitem{cohsc}
H.B.~Li and H.T.~Wong, J. Phys. {\bf G 28}, 1453 (2002).
\end{thebibliography}
\end{document}